\documentclass[aps, prd, letterpaper, amsmath, amssymb,twocolumn,floatfix, nofootinbib]{revtex4-1}
\usepackage[pdftex]{graphicx}
\usepackage[pdftex, pdfstartview={FitH}, pdfnewwindow=true, colorlinks=true, citecolor=blue, filecolor=blue, linkcolor=blue, urlcolor=blue, pdfpagemode=UseNone, bookmarks=false]{hyperref}

\usepackage{comment}
\usepackage{hyperref}   
\usepackage{amsmath}    
\usepackage{graphicx,caption}   
\usepackage{verbatim}   
\usepackage{color}      
\usepackage{float}

\newcommand{\beq}{\begin{equation}}
\newcommand{\eeq}{\end{equation}}
\newcommand{\psib}{\ensuremath{\overline{\psi}}}

\usepackage[T1]{fontenc} 

\begin{document} 
\title{\boldmath Simulations of $ SU(2) $ lattice gauge theory with dynamical  reduced staggered fermions }
\author{Simon Catterall}
\email{smcatter@syr.edu}
\author{Nouman Tariq}
\email{ntbutt@syr.edu}

\affiliation{Department of Physics, Syracuse University, New York 13244, United States}
  \date{\today}

\begin{abstract}We simulate $ SU(2) $ lattice gauge theory using dynamical reduced staggered fermions. The latter lead
to two rather than four Dirac fermions in the continuum limit. We review the derivation and properties of reduced staggered
fermions and show that in the case of fields in the fundamental representation of $SU(2)$ the theory does
not exhibit a sign problem and can be simulated using the RHMC algorithm. We present results on lattices up to $16^4$
for a wide range of bare fermion masses. We find a single site condensate appears at strong coupling that spontaneously breaks the one global $U(1)$ symmetry remaining in the reduced fermion action.
\end{abstract}

\maketitle
\flushbottom

\section{Introduction}
Simulations of gauge theories using staggered fermions have a long history going back to the early days of
lattice gauge theory. In recent years they have allowed for precision studies of hadronic quantities of crucial
importance to experimental efforts to test and constrain the Standard Model ~\cite{Aoki:2016frl}.

As is well known the four dimensional naive staggered fermion action yields not one but
four Dirac fermions in the continuum limit. It is less well appreciated that this replication can be halved by
an additional thinning of lattice degrees of freedom to create what are called {\it reduced} staggered fermions.
At first glance this fact seems
to imply that the reduced fermion would  be a better choice than the usual staggered
fermion for simulations. It was realized early on that this was not the case; for QCD the resulting fermion
determinant is not real, positive definite and furthermore it is not possible to write gauge invariant single site
mass terms in such a theory \cite{vandenDoel:1983mf}. 

In this paper we point out that these problems can be evaded for gauge groups with pseudoreal representations.
As an example we consider the case of fermions transforming in the fundamental representation of $SU(2)$.
Quenched simulations of this model have been studied in ~\cite{Follana:2006zz} but the only work we are aware of with
dynamical fermions was carried out in the context of a four fermion model with off-site Yukawa couplings
\cite{Catterall:2013koa}.
In this paper we study the case of the $SU(2)$ gauge theory with both single site and one-link mass terms. We show that
the corresponding single site fermion condensate dominates at strong coupling in the thermodynamic limit as the fermion masses are sent to
zero. The appearance of a single site condensate 
breaks the one remaining global $U(1)$ symmetry in the reduced fermion action and leads to a light pion which is also measured in our
simulations.
This $U(1)$ symmetry breaking
is consistent with the RMT analysis of ~\cite{Damgaard:2001fg}.

One of our motivations for this work comes from recent studies of a four fermion model built from four reduced staggered fermions.In three dimensions there is good evidence that the weak and strong couplings phases of this theory are separated by a continuous phase transition with non-trivial critical exponents ~\cite{Catterall:2015zua,Ayyar:2014eua}.
Moreover, while no symmetries are broken in the strong coupling phase, the system nevertheless generates a mass gap. Thus the phase transition does not seem to be describable in terms of a Landau-Ginzburg effective theory. 

In four dimensions a very narrow broken symmetry phase reappears between the weak and strong coupling phases ~\cite{Catterall:2016dzf,Ayyar:2016lxq} but there is evidence that this broken symmetry phase may be evaded in an
expanded phase diagram  corresponding to a Higgs-Yukawa generalization of the model \cite{Catterall:2018pzf, Catterall:2017ogi}. This latter model possesses a global $SO(4)=SU(2)\times SU(2)$ symmetry
with the Yukawa interaction coupling the staggered fermions to a scalar field living in the adjoint representation of one of these $SU(2)$s.
A natural extension of this model then replaces the scalar field with an $SU(2)$ gauge field which we conjecture is capable of generating the same
four fermion condensate now via strong gauge interactions. As a first step in this direction we need to understand the phase structure and symmetry
breaking patterns of reduced staggered fermions interacting via a $SU(2)$ gauge field - the study reported here.

\section{Action and symmetries}

For completeness we repeat here the derivation of the reduced staggered fermion action \cite{vandenDoel:1983mf}.
Starting with the full {\it massless} staggered action 
\begin{equation}
\label{eq:fs}
S_F = \sum_{x,\mu}\frac{1}{2}\eta_\mu(x) \psib(x) \left[U_{\mu}(x) \psi(x+\mu) - U_\mu^\dagger (x-\mu) \psi(x-\mu)\right]
\end{equation}
where the staggered fermion phases are given by
\begin{equation}
\eta_\mu(x)=\left(-1\right)^{\sum_{i=1}^{\mu-1}x_i}\end{equation}
we project down to \textit{reduced} staggered variables. 
\begin{equation}
\label{eq:p}
\begin{split}
\psib(x) \to \frac{1+\epsilon(x)}{2} \psib(x) 
\\
\psi(x) \to \frac{1-\epsilon(x)}{2} \psi(x)
\end{split}
\end{equation}
where the parity factor $\epsilon(x)=\left(-1\right)^{\sum_{i=1}^4x_i}$.
Since $ \psib $ is only defined on even sites we can relabel it as $ \psi^T $. Furthermore, we can introduce
a new gauge field $\mathcal{U}_\mu(x)$ defined by
\begin{equation}
\label{eq:x}
\mathcal{U}_{\mu}(x) = \frac{1+\epsilon(x)}{2} U_{\mu}(x) + \frac{1-\epsilon(x)}{2} U^{*}_{\mu}(x) 
\end{equation}
and rewrite the resultant {\it reduced} staggered action in the form
\begin{equation}
S_F = \sum_{x,\mu} \frac{1}{2}\eta_\mu(x)\psi^{T}(x) \mathcal{U}_{\mu}(x) \psi(x+\mu) 
\end{equation}
By taking the transpose of this equation it can be written equivalently as
\begin{equation}
S_F=\sum_{x,\mu}\frac{1}{2}\psi^T(x)\eta_\mu(x)\Delta_\mu(x)\psi(x)\end{equation}
where
\begin{equation}
\Delta_\mu\psi(x)=\frac{1}{2}\left(\mathcal{U}_\mu(x)\psi(x+\mu)-\mathcal{U}_\mu^T(x-\mu)\psi(x-\mu)\right)\end{equation}
which reveals explicitly the antisymmetric character of the reduced fermion operator. This reduced action is
invariant under two symmetries in addition to gauge invariance, a continuous $U(1)$ symmetry which acts on the fermions
\begin{equation}
\psi(x)\to e^{i\alpha\epsilon(x)}\psi(x)\end{equation}
and a discrete shift symmetry 
\begin{equation}
\psi \to \xi_{\rho}\psi(x + \rho)
\end{equation}
where  $ \xi_{\mu} = (-1)^{\sum^{d-1}_{i=\mu+1} x_{i}} $. 
Since for reduced fermions one keeps only $\psi$ or $\overline{\psi}$ at each site the usual
staggered fermion mass term does not exist. However $\psi^a\psi^b\epsilon_{ab}$ is clearly a gauge invariant
fermion bilinear for fermions transforming in the fundamental representation of $SU(2)$ and can hence  be added to the fermion action\footnote{Notice the analog
of this term vanishes for two continuum Weyl fermions because of an additional contraction over Lorentz
indices unless the fermions carry additional flavor indices. }. 
\begin{equation}
\delta S=O_S(x)=m\sum_x \epsilon(x)\psi^a\psi^b\epsilon_{ab}\end{equation}
To understand why the parity factor $\epsilon(x)$ appears in the mass term consider the full fermion operator 
\begin{equation}
D=\eta_\mu(x)\Delta^{ab}_\mu +m\epsilon(x)\epsilon^{ab}
\end{equation}The poles of the associated propagator are determined by the zeroes of $D^2$. 
Using the fact that the parity operator $\epsilon(x)$ anticommutes with the symmetric difference operator $\Delta_\mu$ allows one to write
\begin{equation}
-D^2=-\Delta^\mu\Delta_\mu+m^2\end{equation}
which exhibits the correct pole structure for a massive fermion (in Euclidean space). Notice that this  mass operator induces
the breaking $U(1)\to Z_2$.Alternatively, we can retain the $U(1)$ symmetry by adding a gauge
invariant one link mass term which then breaks the shift symmetry.
\begin{equation}
O_L(x) = m_1\sum_{x,\mu} \frac{1}{2}\xi_{\mu}(x) \epsilon(x) \psi^T(x) \mathcal{M}_\mu\psi(x)\end{equation}
where \begin{equation}
\mathcal{M}_\mu\psi(x)=\frac{1}{2}\left[ \mathcal{U}_{\mu}(x)\psi^b(x+\mu)+
\mathcal{U}^T_{\mu}(x-\mu) \psi(x-\mu) \right]
\end{equation}
Notice the addition of $O_S$ and $O_L$ to the action preserves the antisymmetry of the fermion operator.
In our numerical work we have investigated the effects of both of these mass terms.
For a full staggered field the symmetry breaking patterns are a little different. For such a staggered field in a pseudoreal representation we can pair the $\psi$ and $\psib$
at each site into a doublet with the kinetic operator now being invariant under a $U(2)$ symmetry. 
In this case a site mass term now breaks $U(2) \to O(2)$. Such a symmetry breaking pattern could also be obtained by using two reduced staggered fields.
\begin{figure}[tbp]
\includegraphics[width=.5\textwidth]{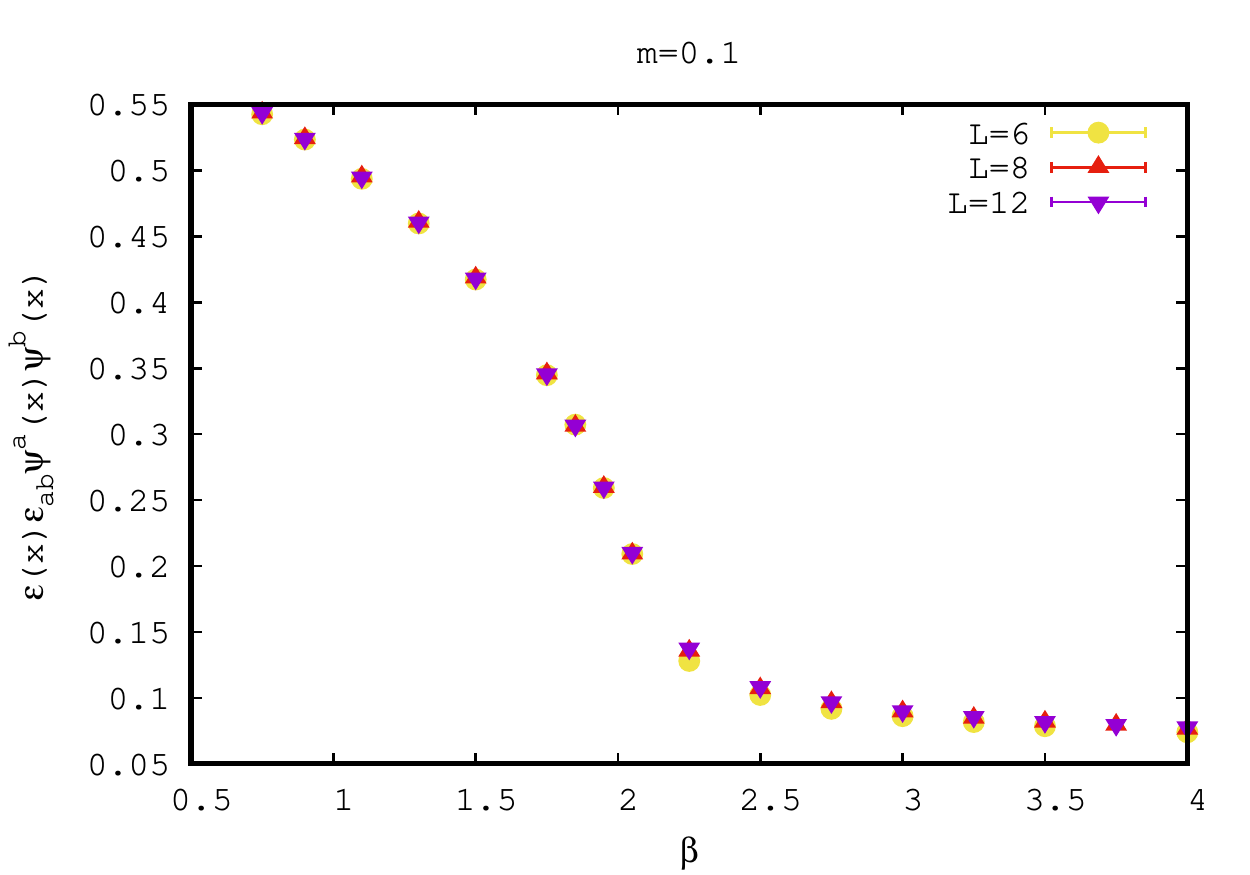}
\caption{\label{fig:bi} $ \langle O_S(x) \rangle  $ } 
\includegraphics[width=.5\textwidth]{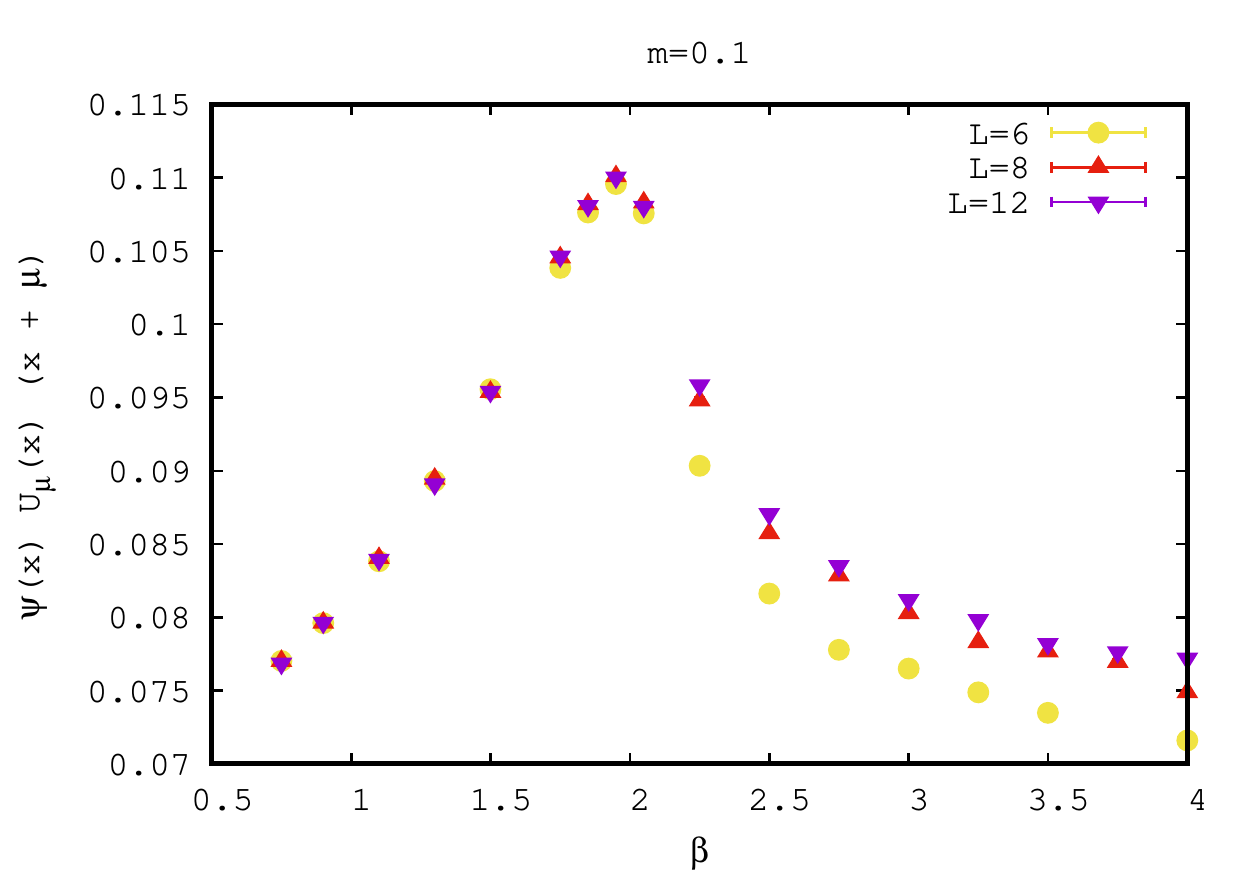}
\caption{\label{fig:link} $ \langle O_L(x) \rangle  $ } 
\end{figure}
Once we integrate the fermions we generate a Pfaffian.Since the fermion operator is antisymmetric
its eigenvalues come in pairs $(\lambda,-\lambda)$. Additionally the pseudo-real nature of the representation
implies that $U_\mu^*(x)=\sigma_2U_\mu(x)\sigma_2$ which ensures that every eigenvalue $\lambda_n$ (generically complex) and corresponding eigenvector $v_n$ is paired with another with eigenvalue $\lambda_n^*$ and eigenvector $\sigma_2v_n^*$.
This quartic structure of the spectrum ensures that the Pfaffian is positive definite and can hence be written $
Pf(D) = {\rm det}( D^{\dagger} D) ^{\frac{1}{4}}$ which is suitable for use in a Monte Carlo algorithm~\cite{Clark:2003na}\footnote{An exception to this can occur if the fermion operator develops a purely real eigenvalue which is then unpaired. We have seen no sign of such eigenvalues in our simulations.}.
For the gauge part of the $ SU(2) $ action we employ the standard Wilson action 
\begin{equation}
S_{G} = \sum_{x}\sum_{\mu < \nu} -\frac{\beta}{2N} Tr [ U_{\mu\nu }(x) + U^{\dagger}_{\mu\nu}(x) ]
\end{equation}
\section{Numerical Results}
\begin{figure}[tbp]
\includegraphics[width=.5\textwidth]{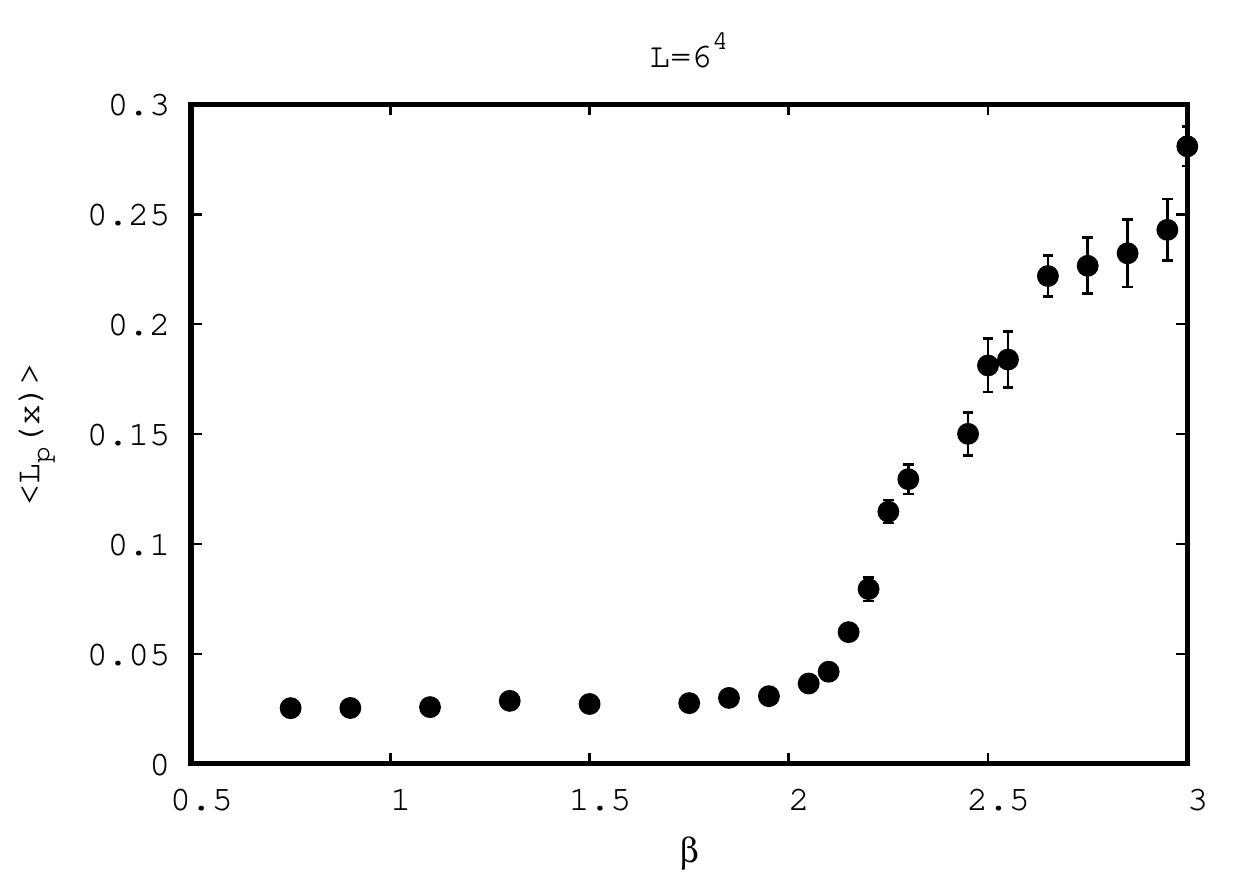}
\caption{\label{fig:line} $ \langle L_{p}(x) \rangle $ } 
\end{figure}

\begin{figure}[tbp]
\includegraphics[width=.5\textwidth]{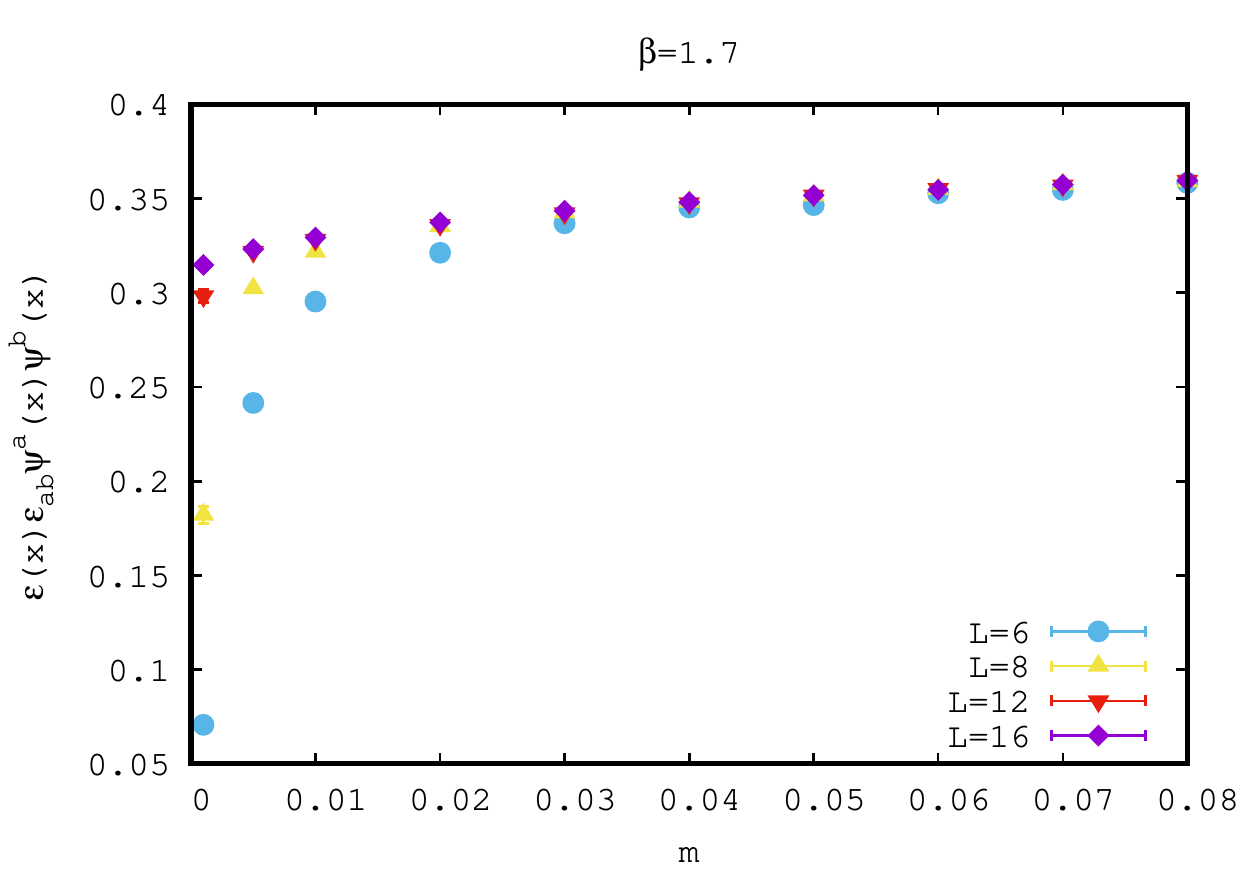}
\caption{\label{fig:bivsm1.7} $ \langle O_S(x) \rangle $ } \par
\includegraphics[width=.5\textwidth]{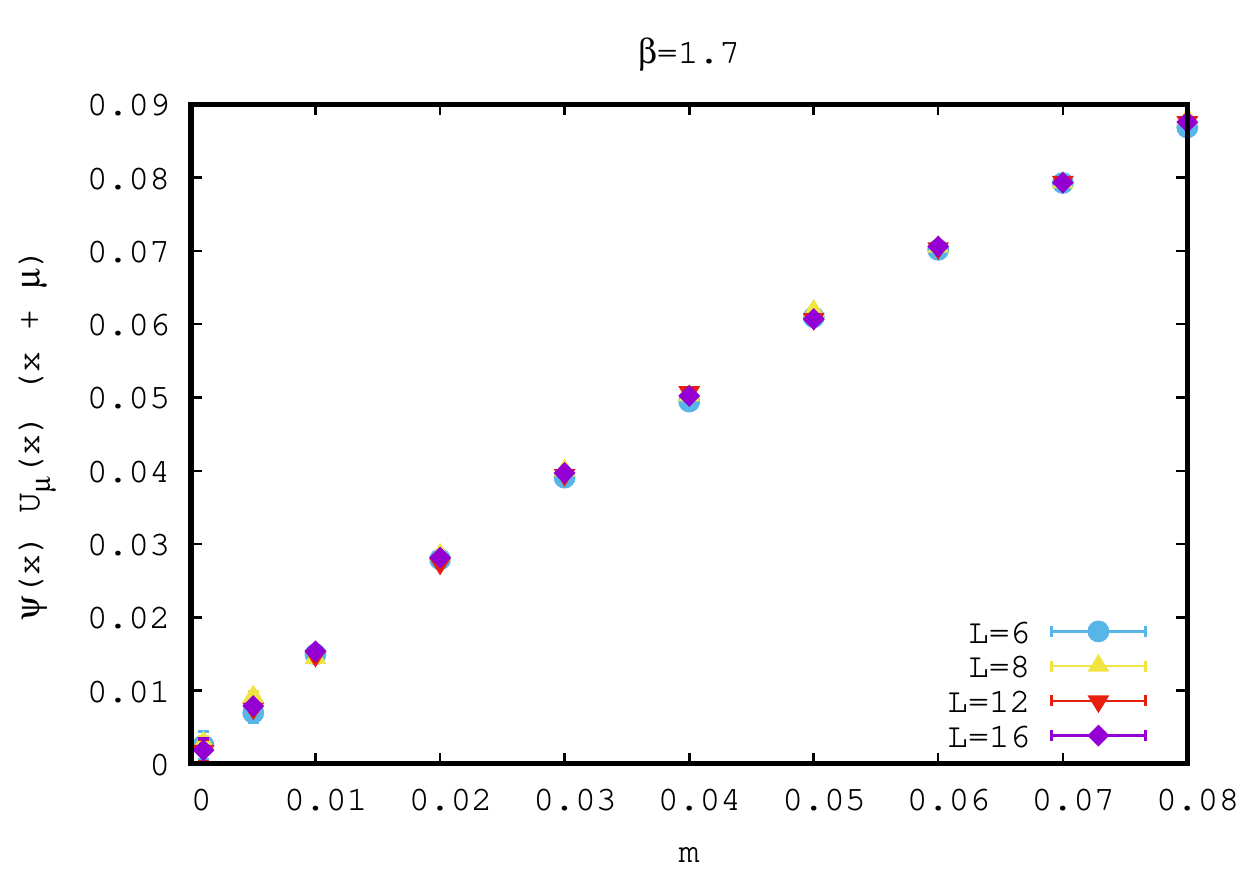}
\caption{\label{fig:linkvsm1.7} $ \langle O_L(x) \rangle  $ } \par
\end{figure}

\begin{figure}[tbp]
\includegraphics[width=.5\textwidth]{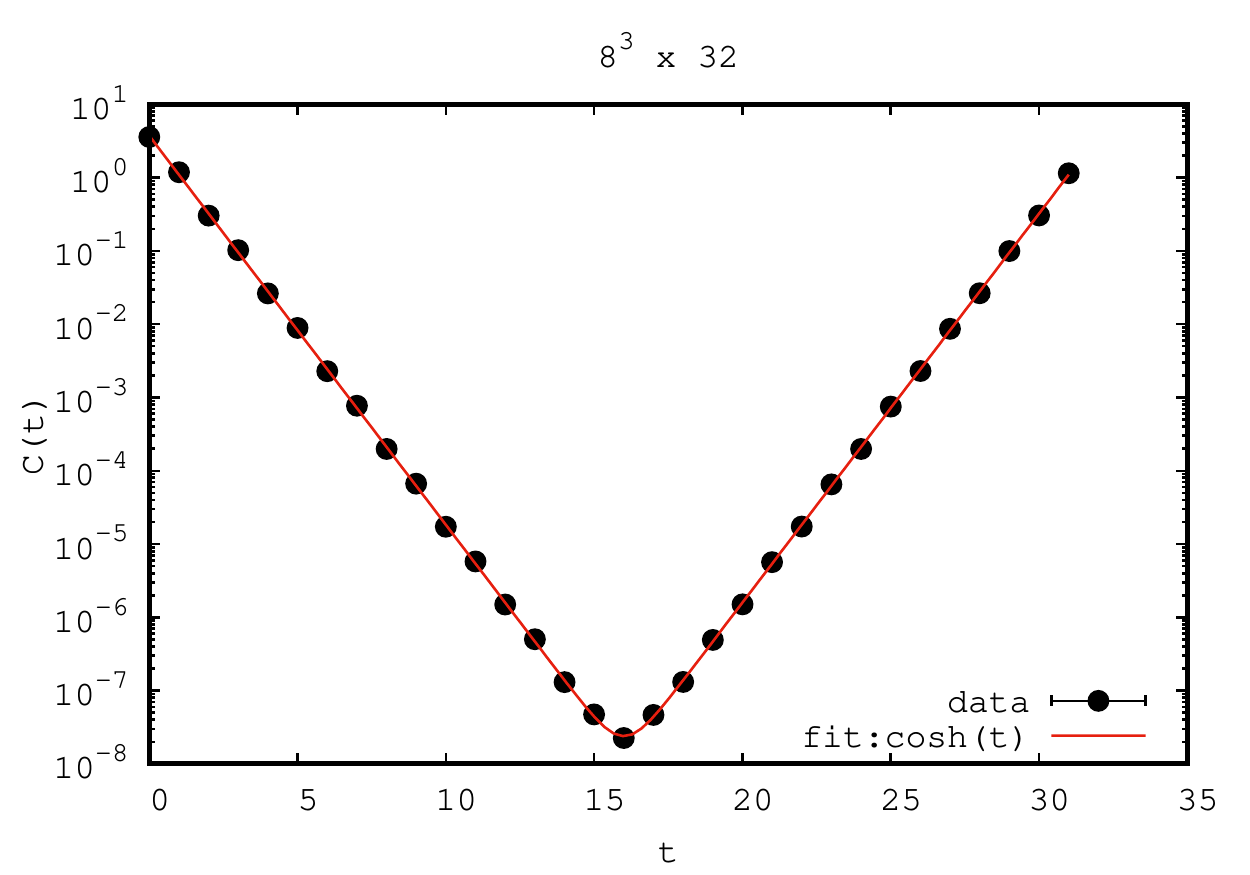}
\caption{ $ \langle C(t) \rangle_{ m_{q}=0.2}  $ } 
\label{fig:corr_0.2}
 \includegraphics[width=.5\textwidth]{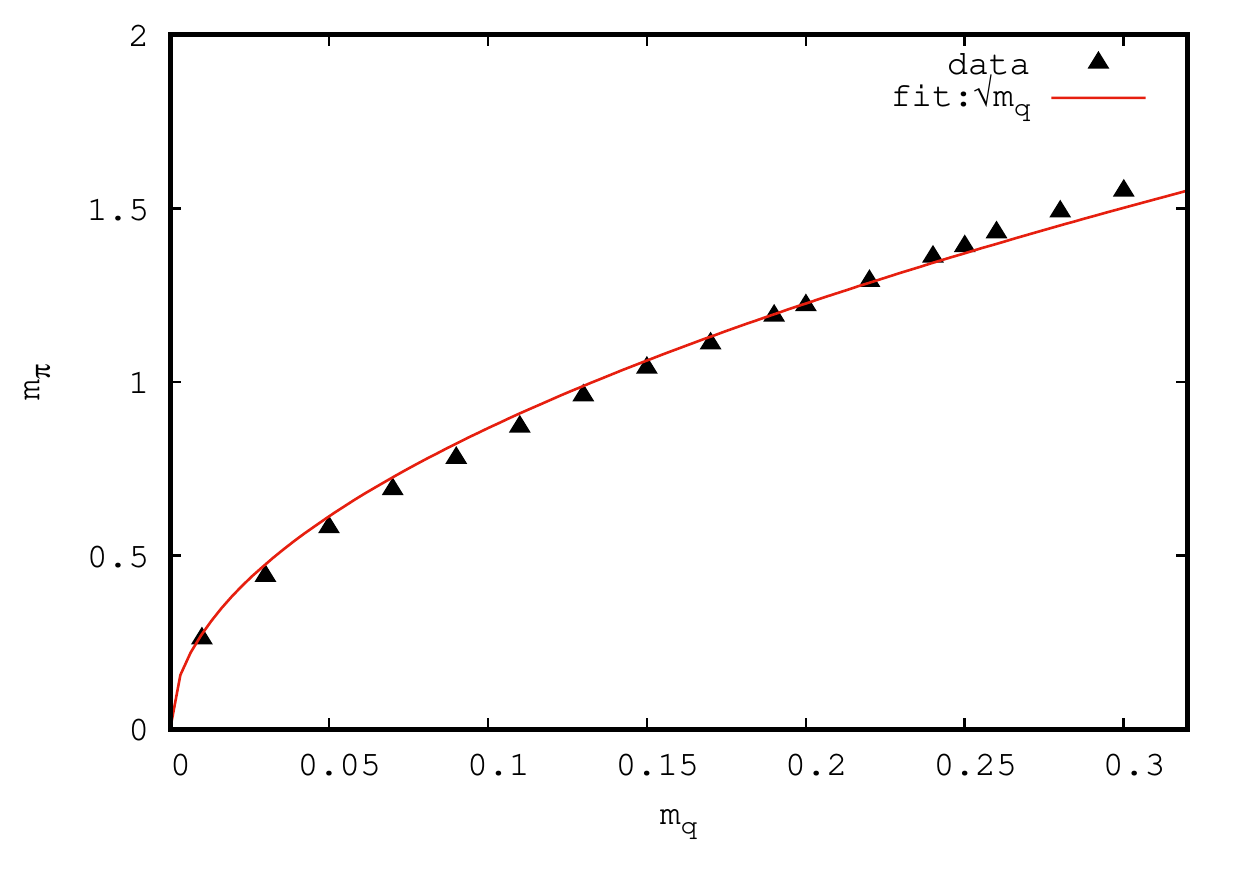}
\caption{$ m_{\pi}  $ vs $ m_{q} $ }
\label{fig:pi_mass}
\end{figure}

We implemented the RHMC algorithm to simulate the model exploring lattice volumes up to $ 16^4 $ with gauge couplings spanning $\beta=0.5-4.0$ and for a wide range of
site and link masses. Fig.~\ref{fig:bi} and fig.~\ref{fig:link} show plots of the expectation values of the site and link bilinears for $m=m_1=0.1$ as a function of the gauge coupling $\beta$ for several
lattice volumes. Both vevs are driven to small values for large $\beta$ as expected since the model enters s deconfined phase in that regime which can be seen clearly in fig.~\ref{fig:line} which shows the Polyakov line as a function of gauge coupling for $m=m_1=0.1$ on a $6^4$ site lattice. The Polyakov line functions as a quasi-order parameter for the  breaking of center symmetry and runs from  small to large values as the system deconfines.Of course the key question is whether one or more of these bilinear vevs remains nonzero in the thermodynamic limit as the bare fermion mass is sent to zero. We focus on the largest values of the (inverse) gauge coupling (smallest lattice spacing) which clearly lie within the confining regime of the theory on the lattice volumes we have simulated. 
\begin{figure}[tbp]
\includegraphics[width=.5\textwidth]{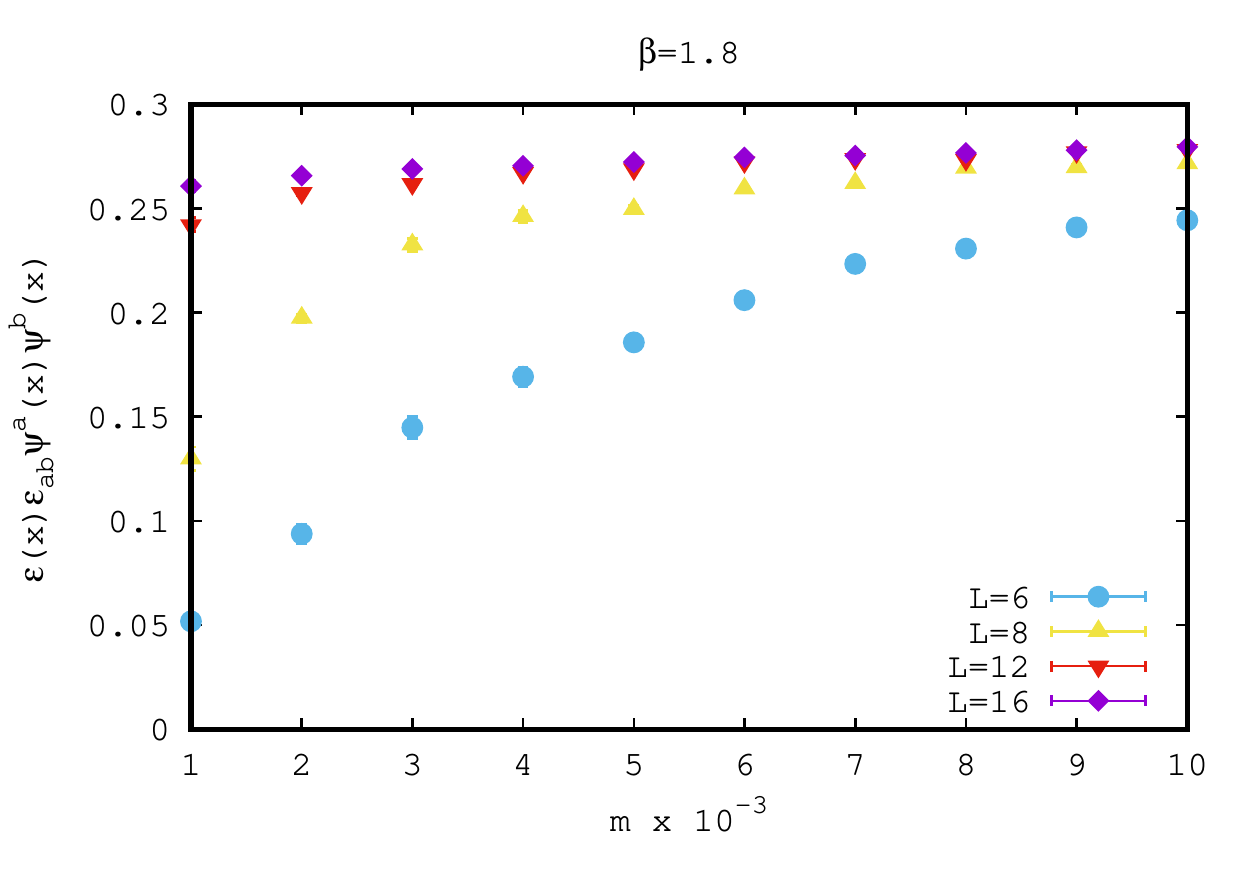}
\caption{\label{fig:bivsm} $ \langle O_S(x) \rangle $ } 
\includegraphics[width=.5\textwidth]{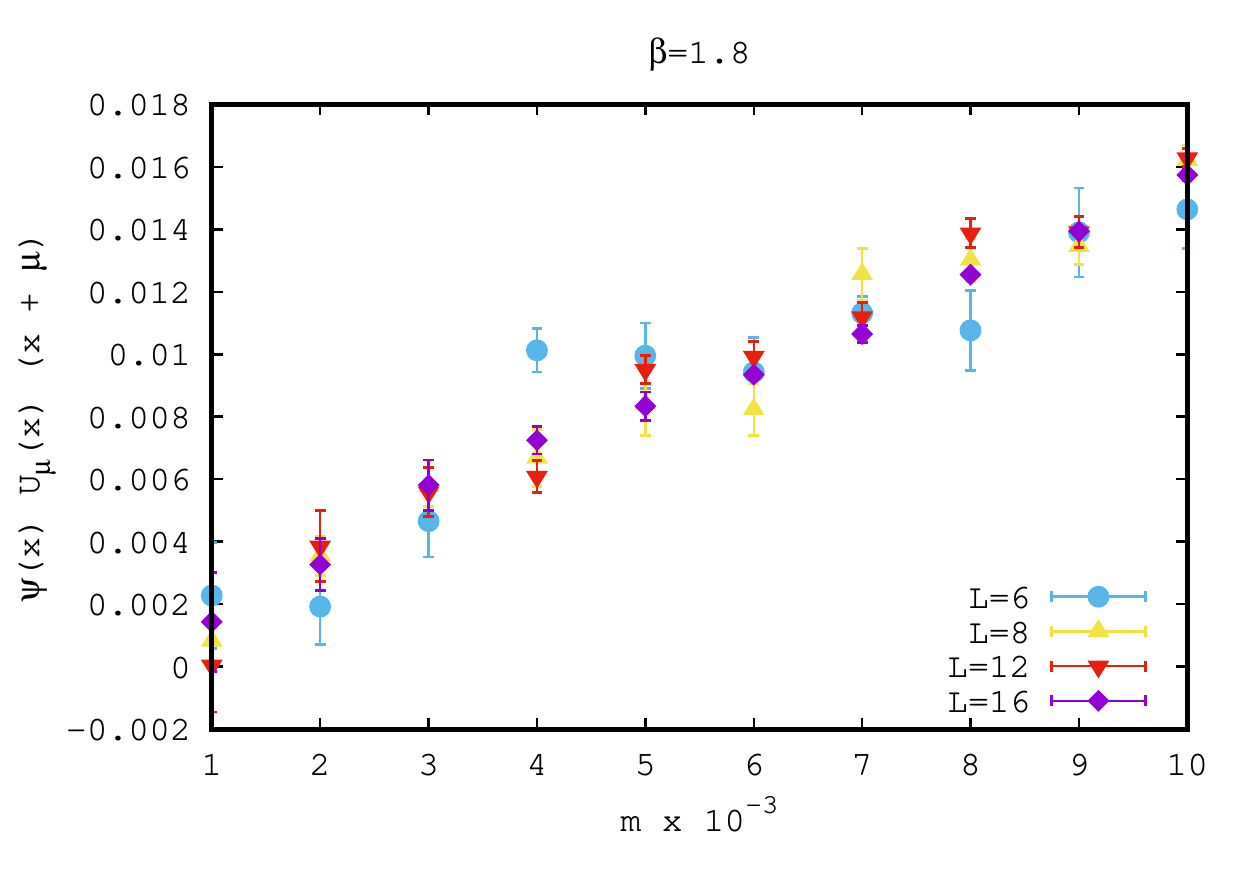}
\caption{\label{fig:linkvsm} $ \langle O_L(x) \rangle  $ } 
\end{figure}
In fig.~\ref{fig:bivsm} and fig.~\ref{fig:linkvsm} we show plots of the expectation values of the two blinears versus the bare fermion mass $m=m_1$ for gauge coupling  $\beta=1.8$ for a range of lattice volumes. Clearly the link vev shows no strong volume dependence and smoothly goes to zero as the external mass is sent to zero. This is consistent with work by Follana et al ~\cite{Follana:2006zz} for full staggered fermions in quenched approximation.The site bilindear shows a very different behavior with the measured vev growing with volume at small mass.This is the behavior needed for a non-zero condensate to survive the
zero mass limit and indeed the data is quite consistent with the presence of a non-zero site condensate in that limit.
To gain confidence in this result we repeated the analysis for $\beta=1.7$ (fig.~\ref{fig:bivsm1.7} and fig.~\ref{fig:linkvsm1.7}) corresponding to a larger value of the lattice spacing. The overall conclusion remains the same and we infer that the preferred breaking channel for the simple reduced staggered fermions studied here corresponds to  $U(1)\to Z_2$\footnote{Similar results were observed by Follana~\cite{Follana:2006zz} in the quenched approximation although the nature of the condensate changed when a smeared action was employed.}
We can confirm these conclusions by looking for the corresponding Goldstone boson - the pion - whose correlator is given by
\begin{equation}
\langle \phi(x) \phi(y) \rangle = \langle \epsilon^{ab} \psi^a(x)\psi^b(x) \epsilon^{cd} \psi^c(y) \psi^d(y) \rangle
\end{equation}
A typical correlator is shown at $m=0.1$ and $\beta=1.8$ on a $8^3\times 32$ lattice in Fig.\ref{fig:corr_0.2}.
We use the standard fit  $ C_{\pi} (t) \sim  A\left[\exp( -am_{\pi} (t) + \exp(-am_{\pi}(T-t)) \right]$ to extract the pion mass. 
In Fig.\ref{fig:pi_mass} we plot the pion mass as a function of the bare quark mass.  The solid line is a fit to the expected square root form and corresponds to the standard GMOR prediction confirming that this state is indeed a pion resulting from spontaneous breaking of the $U(1)$ symmetry.
\section{Summary}
In this paper we perform the first studies of $SU(2)$ lattice gauge theory
with dynamical reduced staggered fermions.The pseudoreal nature of the fundamental representation of $ SU(2) $ allows us
to employ the standard RHMC algorithm without encountering a
sign problem. Unlike $SU(N)$ for $N>2$ a gauge invariant site mass term is allowed 
and we investigate the model including both this term and
a gauge invariant one-link mass operator. We find strong evidence that a site bilinear fermion condensate is formed at strong coupling spontaneously breaking an exact $U(1)$ symmetry down to $Z_2$. We find good evidence for the
corresponding Goldstone boson - the pion. These results are
consistent with previous studies that used the spectrum of low-lying eigenmodes of the quenched Dirac operator to find evidence for chiral symmetry breaking in
this theory. Our results strengthen these conclusions and support the analysis given in \cite{Damgaard:2001fg}. 
This work is motivated by an attempt to understand some of the novel phase structure in a related Higgs-Yukawa model involving reduced staggered fermions interacting with $SU(2)$ gauge fields.The current work establishes the bedrock for understanding the results of those studies the results of which will be reported soon.
\acknowledgments
This work is supported in part by the U.S. Department of Energy,Office of Science, Office of High Energy Physics,under Award Number DE-SC0009998. Numerical computations were performed at Fermilab using USQCD resources.
\bibliography{SU2}
\bibliographystyle{plain}

\end{document}